\documentclass [aps,
superscriptaddress,
reprint,
amsmath,amssymb,
showkeys,
prl,floatfix,longbibliography,
]{revtex4-1}

\usepackage{graphicx}
\usepackage{dcolumn}
\usepackage{bm}
\usepackage{color}
\usepackage{float}
\usepackage[utf8]{inputenc}
\usepackage[mathlines]{lineno}
\usepackage{multirow}
\usepackage{hyperref}
\usepackage{amsmath}
\usepackage{xr}
\usepackage{soul,xcolor}
\soulregister\cite7
\soulregister\ref7
\setstcolor{blue}
\newcommand{\itO}{\textit{o}-}
\newcommand{\itT}{\textit{t}-}
\newcommand{\itC}{\textit{c}-}

\newcommand{\wn}{cm$^{-1}$}
\newcommand{\wnws}{cm$^{-1}$ }

\begin{document}

	\title{The tetragonal phase of CH$_{3}$NH$_{3}$PbI$_{3}$ is strongly anharmonic}
	\author{Rituraj Sharma}
	\thanks{These authors contributed equally}
	\affiliation{Department of Materials and Interfaces, Weizmann Institute of Science, Rehovot 76100, Israel}
	\author{Zhenbang Dai}
	\thanks{These authors contributed equally}
	\affiliation{Department of Chemistry, University of Pennsylvania, Philadelphia, Pennsylvania 19104--6323, USA}
	\author{Lingyuan Gao}
	\affiliation{Department of Chemistry, University of Pennsylvania, Philadelphia, Pennsylvania 19104--6323, USA}
	\author{Thomas M. Brenner}
	\affiliation{Department of Materials and Interfaces, Weizmann Institute of Science, Rehovot 76100, Israel}
	\author{Lena Yadgarov}
	\affiliation{Department of Materials and Interfaces, Weizmann Institute of Science, Rehovot 76100, Israel}
	\author{Jiahao Zhang}
	\affiliation{Department of Chemistry, University of Pennsylvania, Philadelphia, Pennsylvania 19104--6323, USA}
	\author{Yevgeny Rakita}
	\affiliation{Department of Materials and Interfaces, Weizmann Institute of Science, Rehovot 76100, Israel}
	\author{Roman Korobko}
	\affiliation{Department of Materials and Interfaces, Weizmann Institute of Science, Rehovot 76100, Israel}
	\author{Andrew M. Rappe}
	\email{rappe@sas.upenn.edu} 
	\affiliation{Department of Chemistry, University of Pennsylvania, Philadelphia, Pennsylvania 19104--6323, USA}
	\author{Omer Yaffe}
	\email{omer.yaffe@weizmann.ac.il}
	\affiliation{Department of Materials and Interfaces, Weizmann Institute of Science, Rehovot 76100, Israel}

	\date{\today}
	
	\begin{abstract}
		
Halide perovskite (HP) semiconductors exhibit unique strong coupling between the electronic and structural dynamics. 
The high-temperature cubic phase of HPs is known to be entropically stabilized, with imaginary frequencies in the calculated phonon dispersion relation.
Similar calculations, based on the static average crystal structure, predict a stable tetragonal phase with no imaginary modes.
This work shows that in contrast to standard theory predictions, the room-temperature tetragonal phase of CH$ _{3} $NH$ _{3} $PbI$ _{3}$ is strongly anharmonic.
We use Raman polarization-orientation (PO) measurements and \textit{ab initio} molecular dynamics (AIMD) to investigate the origin and temperature evolution of the strong structural anharmonicity throughout the tetragonal phase.
Raman PO measurements reveal a new spectral feature that resembles a soft mode. 
This mode shows an unusual continuous increase in damping with temperature which is indicative of an anharmonic potential surface.
The analysis of AIMD trajectories identifies two major sources of anharmonicity: the orientational unlocking of the [CH$ _{3} $NH$ _{3}$]$^+$ ions and large amplitude octahedral tilting that continuously increases with temperature.
Our work suggests that the standard phonon picture cannot describe the structural dynamics of tetragonal CH$ _{3} $NH$ _{3} $PbI$ _{3}$.
	\end{abstract}

	\maketitle
	
	Methylammonium lead iodide (MAPI) has emerged as an outstanding absorber material for photovoltaics~\cite{Brenner2016,Gratzel2017,Jeon2014}. 
	Recent evidence indicates that the beneficial optical and electronic properties of halide perovskites (HPs) are intimately associated with their unusual structural dynamics and mechanical softness~\cite{Egger2018}.
	The cubic phase of HPs was shown to exhibit strongly anharmonic thermal fluctuations with light scattering properties that resemble those of liquids~\cite{Poglitsch1987,Mayers2018,Yaffe2017,Whalley2017}. 
	Yet, little is known about the structural dynamics of the room-temperature tetragonal phase of MAPI. 

%

MAPI crystallizes into an \textit{ABX${}_{3}$} type perovskite structure, with PbI${}_{6}$ in a corner-sharing octahedral network, and MA molecules inside the cuboctahedral cavities.\\
It shows two structural phase transitions:\\ 
orthorhombic~(\itO)~to~tetragonal~(\itT) at $\approx$~162~K, and tetragonal to cubic (\itC) at $\approx$~327 K~\cite{Poglitsch1987,Whitfield2016}.
Similar to other perovskite crystals, the calculated harmonic phonon dispersion relation of the cubic phase of MAPI shows imaginary frequencies~\cite{Beecher2016,Brivio2015}, which indicates that the cubic structure has double-wells in its potential-energy surface and thus, is entropically stabilized (i.e. strongly anharmonic)~\cite{Beecher2016,Zhu2019a}.
On the other hand, the orthorhombic phase does not have any imaginary frequency modes, consistent with current expectations that MAPI exhibits mostly harmonic behavior in the orthorhombic phase~\cite{Sharma2019}.
	
Interestingly, the phonon dispersion relation of the \itT phase of MAPI does not exhibit imaginary frequencies~\cite{Wang2016,Brivio2015}, implying the stability of the tetragonal structure of MAPI. Nevertheless, many seemingly contradictory characteristics that prevail at room temperature in the \itT phase, such as low mobility but high carrier lifetime, small Urbach energy but high density of structural defects~\cite{DeWolf2014,Walsh2016,Egger2018} cannot be explained in a harmonic picture but point toward  strong structural anharmonicity also in the \itT phase. 
	
In this work, we investigate the complex structural dynamics of the \itT phase of MAPI. 
Using temperature-dependent polarization orientation (PO) Raman measurements, we discover new spectral features and trends that indicate that the \itT phase of MAPI is strongly anharmonic.
A definitive assignment of the Raman spectral features to real-space atomic motions (beyond normal-mode analysis) is obtained using \textit{ab initio} molecular dynamics (AIMD). 
We employed this real-space analysis to elucidate the microscopic mechanisms governing the strongly anharmonic structural dynamics in \itT-MAPI.



	\begin{figure*}
	\includegraphics[width=17 cm, keepaspectratio=true]{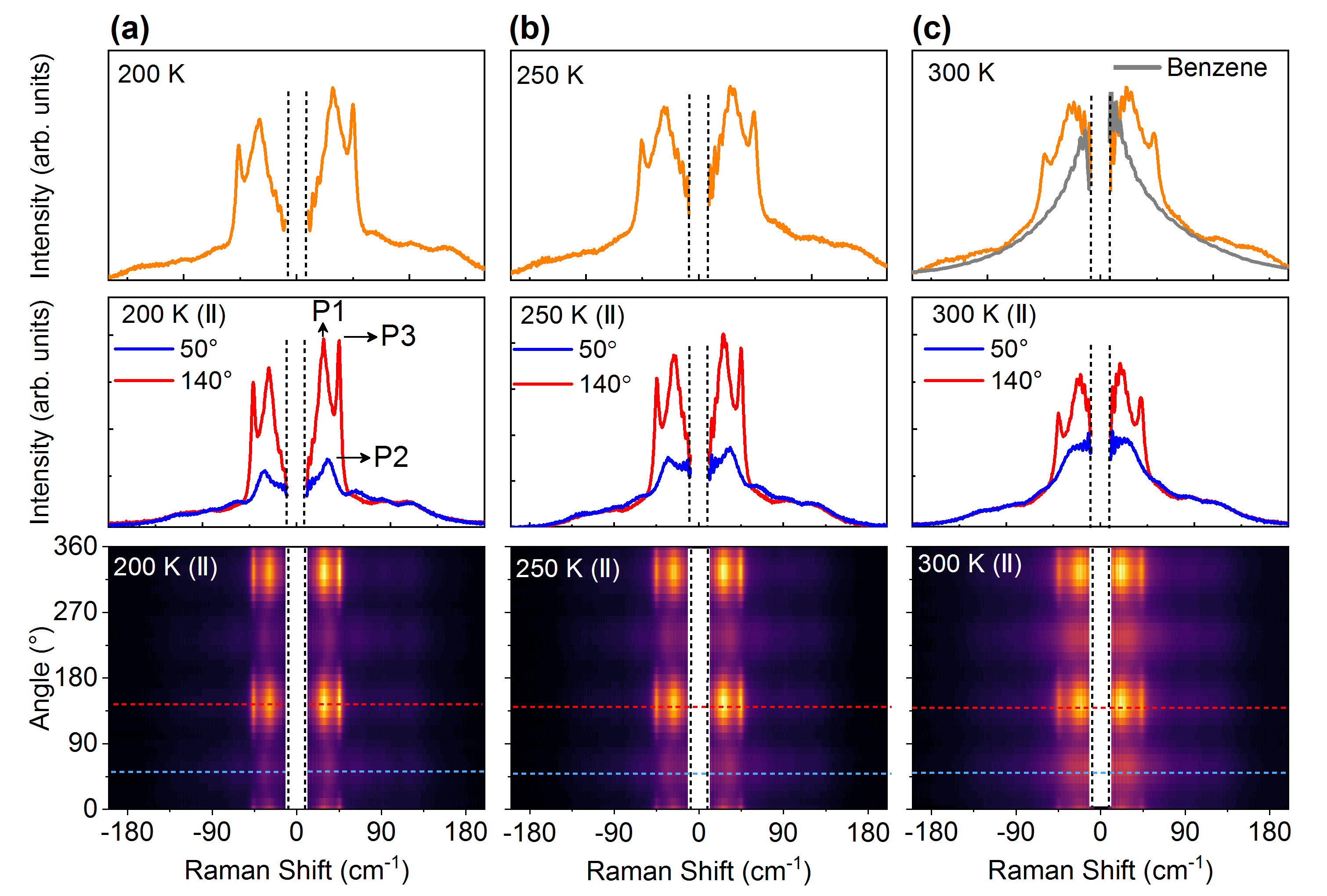}
	\caption{Temperature evolution of PO-dependent Raman spectra of MAPI measured at 200 K (a), 250 K (b) and 300 K (c). The upper panel shows the unpolarized spectra (integrated over all polarization angles). The spectrum of MAPI at 300 K is compared with that of Benzene in (c). Contour plots for PO dependence of the Raman spectra in the parallel configuration are shown in bottom panel. Cross-sections of the contours corresponding to the angular positions shown by dotted blue and red lines are shown in the middle panel (blue and red curves). The 0$^\circ$ polarization angle is arbitrarily defined and is identical for all measurements. 
	}
	\label{PO_Raman}	
\end{figure*}

    We perform low-frequency Raman measurements on single crystals oriented along the [110] direction relative to the Bravais lattice unit cell. X-ray diffraction results are shown in Fig.\ S1 of the supplemental material (SM)~\cite{SM}.
    Sub-band gap excitation at 1064~nm (1.16~eV) is used to overcome the damage related to the light absorption and poor thermal conductivity of MAPI that occurs with shorter wavelength excitation.
     Figure~\ref{PO_Raman} (top panel) shows the unpolarized Raman spectra at three different temperatures throughout the \itT phase, displaying both anti-Stokes and Stokes scattering. 
    The low-temperature (200~K) spectrum shows two distinct peaks at $ \approx$~28 and 45~\wn. As the temperature increases, the spectrum broadens and red shifts. 
At 300~K, a broad spectrum with a peak near 45~cm$^{-1} $ is observed. Interestingly, the room temperature spectrum of MAPI resembles the Raman spectrum of a fluid.
 In the upper panel of Fig.~\ref{PO_Raman}c, we have compared the room-temperature spectra of MAPI with benzene. The extraordinary similarity between these two spectra indicates that despite the crystalline order of the tetragonal structure, MAPI exhibits liquid-like, relaxational motions~\cite{Perrot1981,Nielsen1979}.
    
    Next, we use PO Raman spectroscopy (Fig.~\ref{PO_Raman}, lower panels) to investigate the origin of structural anharmonicity in \itT phase MAPI. The details of the experiment are discussed in the SM~\cite{SM}. Briefly, the crystal oriented along the [110] direction is excited by plane-polarized laser light with polarization $e_i$. The scattered light is then filtered by another polarizer (analyzer) oriented parallel or perpendicular to the incident light. This PO measurement is repeated after each incremental rotation of the polarization of the incident light. 

    The contour plots in the lower panels of Fig.~\ref{PO_Raman} show the angular dependence of polarized Raman spectra for measurements performed at 200~K, 250~K, and 300~K covering the stability range of the tetragonal phase, in the parallel configuration. It is worth mentioning that the data in Fig.~\ref{PO_Raman} are presented in their raw form without any normalization or baseline correction, and the results are completely reversible with temperature. These plots show a periodic modulation of the intensities of all the Raman peaks with polarization angle.
    The data for the cross configuration are given in the SM~\cite{SM}, Fig.\ S2. The strong and periodic oscillations in intensity are indicative of the long-range order of the crystal.
    In standard quasi-harmonic crystals (e.g. silicon, GaAs), the combination of parallel and perpendicular data sets enables the extraction of the symmetry of each observed mode~\cite{Mizoguchi1989,Hayes2004}. We have recently demonstrated this for the \itO phase of MAPI~\cite{Sharma2019}. However, since the Raman features of the \itT phase are extremely broad and cannot be assigned to specific normal modes, the interpretation of the PO data is not as straightforward.

	\begin{figure}
	\includegraphics[width=8.5cm, keepaspectratio=true]{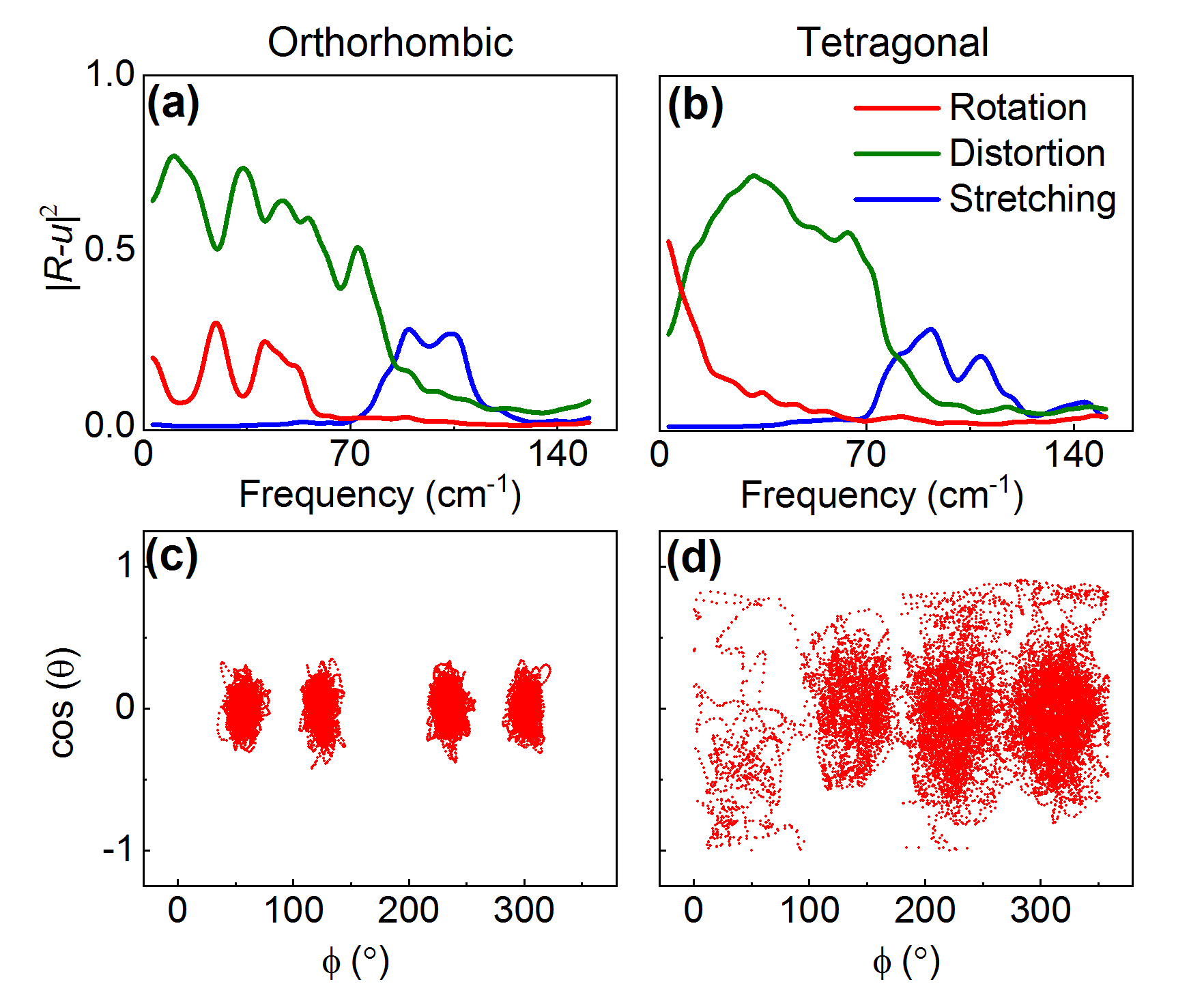}
	\caption{The projections of frequency-filtered AIMD for the orthorhombic (a) and tetragonal (b) phases.  The nearly harmonic low-frequency peaks in (a) the \itO phase  representing rotations of the octahedra (red curve) are replaced (b) by a broad central feature in the \itT phase. Statistics of MA molecular orientations in the (c) \itO and (d) \itT phases. $\phi$ is the azimuthal angle and $\theta$ is the polar angle. The molecule is allowed to visit more orientations when it goes from the \itO~to the \itT phase, indicating the onset of the molecular disorder and strong anharmonic mode couplings.}
	\label{Mode_trajectory}			
\end{figure}

  The low-frequency ($<$50~\wn) spectra show modulation with 180$^\circ$ period in the PO dependence, resulting in two distinct line shapes at 50$^\circ$ and 140$^\circ$ PO angles, denoted by blue and red dotted lines, respectively in Fig.~\ref{PO_Raman} (bottom panel). 
The spectra were deconvoluted using a multi-Lorentz oscillator model (details of the fit procedure are given in the SM~\cite{SM}). We identify six distinct spectral features (labeled as P1-P6). The corresponding widths and positions extracted from the fits are given in Table S1 and Figs. S3 and S4 in the SM~\cite{SM}. 
In Fig.~\ref{PO_Raman} (middle panel), we observe that all the spectral features of the 140$^\circ$ lineshape appear in the unpolarized spectrum also (top panel).
However, the $\approx$~30~\wnws feature (P2, blue curve) is undetectable in the unpolarized spectrum(top panel of Fig.~\ref{PO_Raman}). 
In that sense, it is a hidden feature. 

Notably, both the relative intensity and width of P2 rapidly increase with increasing temperature (middle panel of Fig.~\ref{PO_Raman}). 
When P2 is completely overdamped (at $\approx$340~K) a \itT\itC phase transition occurs. 
In cubic phase, the 140$^\circ$ and 50$^\circ$ lineshapes become indistinguishable and identical to the unpolarized lineshape (See Fig.~S5 in SM~\cite{SM}).
Therefore, P2 resembles a soft mode that leads to the  \itT\itC phase transition~\cite{Burns1970}.

According to our fitted line width (See SM~\cite{SM}, Table S1), the lifetime of the Lorentz oscillator P2 becomes extremely short ($\approx$~154~fs) at room temperature, at least two orders of magnitude less than the lifetime of the standard optical phonon of crystalline silicon~\cite{Parker1967}. Using the average phonon group velocities reported elsewhere~\cite{Elbaz2017,Qian2016}, we found that our calculated phonon mean free path is $\approx$2.2~\AA, which is shorter than the \textit{t-}MAPI lattice constant (a = 8.8~\AA, c = 12.7~\AA)~\cite{Whitfield2016}.	
This suggests that the standard phonon picture cannot describe the structural dynamics of the crystal. The breakdown of the phonon picture also implies that the Fr\"{o}hlich interaction mechanism, which is considered to be the dominant mechanism for electron scattering in MAPI~\cite{Wright2016a}, is oversimplified. 

To gain microscopic understanding of the complex structural dynamics of the \itT phase of MAPI, we performed AIMD simulations.
First, we use the AIMD trajectories to reproduce the experimental Raman spectra for all three phases of MAPI (Fig. S6 in SM~\cite{SM}).
Next, we assign atomic motions to the spectral features observed in Fig.~\ref{PO_Raman}.
We defined specific real-space ionic motions as shown in Figs. S7, S8, S9 in the SM~\cite{SM}.
AIMD trajectories filtered for a narrow frequency window (0.5 cm$ ^{-1} $) were then projected onto these predefined modes in order to construct a spectrally-resolved decomposition of the ionic motions.
We assigned the spectral feature P2 to a symmetric distortion in the {pseudo-cubic} \textit{xz} and \textit{yz} plane, while the P3 feature is the analogous symmetric distortion in the \textit{xy} plane, as shown in Fig. S10 of the SM~\cite{SM}.
Both P2 and P3 reflect the same distortion pattern but in different planes. 
Evidently, the strong damping with temperature of P2 (middle panel of Fig.~\ref{PO_Raman})  is indicative of strong anharmonicity in the symmetric distortion of pseudo-cubic \textit{xz} and \textit{yz} planes.

	\begin{figure*}
		\includegraphics[width=17 cm]{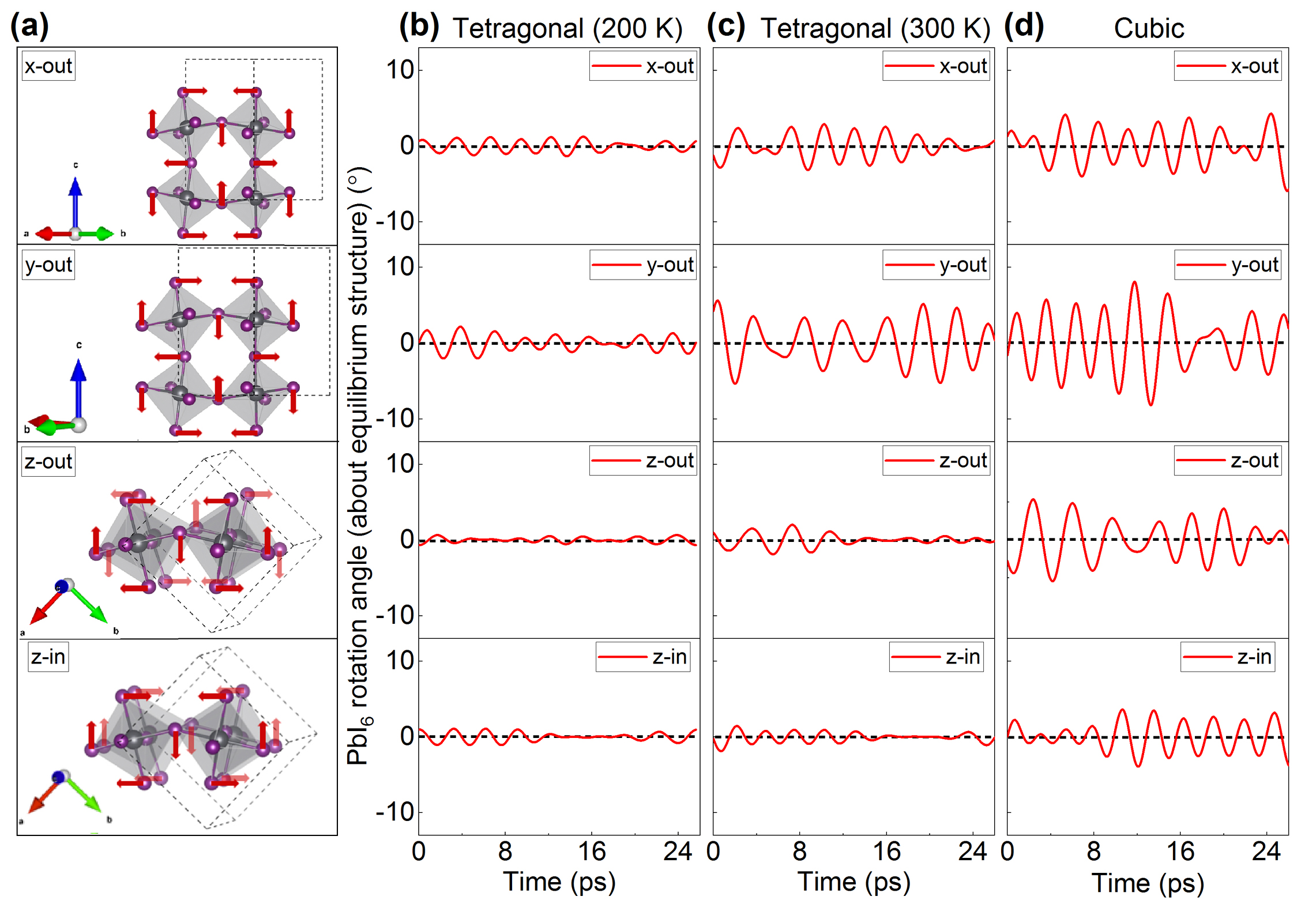}
		\caption{Contrasting the PbI$_6$ rotational motions in the \itT and \itC phases. (a) Illustrations of real-space PbI$_6$ rotational motions.
			(b-d) Time evolution of PbI$_6$ rotation amplitudes in the \itT phase at 200~K (b), 300~K (c), and the \itC phase (d).
			The legend  reports the rotation axis and the relative rotation direction.
			For example, \textit{x}-out represents the out-of-phase rotation along the \textit{x} axis, and \textit{z}-in represents the in-phase rotation along the \textit{z}-axis.
			These amplitudes are obtained from MD trajectories frequency-filtered to 1~\wn.
			The black dashed lines at 0$^\circ$  represent the equilibrium tilting angle of the octahedra around each axis, which has been offset to the 0$^\circ$ for clearer comparison. The equilibrium angle can be found in Table~ S2.
		}
		
		\label{Real_Space}		
		
	\end{figure*}
	 

 
We now elucidate the structural anharmonicity that is related to the \itO\itT phase transition in MAPI.
To do so,  we examine the spectrally-resolved decomposition of the ionic motions. The weight of each mode as a function of frequency is displayed for the \itO and \itT phases in Fig.~\ref{Mode_trajectory}.
The data clearly show that only octahedral distortions and rotations are dominant in the frequency region below 50~\wn. For the \itO phase shown in Fig.~\ref{Mode_trajectory}(a), two distinct peaks corresponding to PbI$_6$ rotational vibrations are found, and an ultra-low-frequency, octahedral rotational motion (5~\wn) is also present. This is in line with our previous observations that the \itO phase is quasi-harmonic and has well defined normal modes~\cite{Sharma2019}. After phase transition to the \itT phase Fig.~\ref{Mode_trajectory}(b), PbI$_6$ rotations (red) change dramatically. The well-resolved features of the \itO phase are replaced by a dominant broad feature that is centered around 0~\wnws (a ``central peak") in the \itT phase. This indicates that the quasi-harmonic PbI$_6$ vibrations of the \itO phase have evolved to relaxational rotations in the \itT phase~\cite{Uchino1996,Gorkovaya2019}.
    The abrupt change in Raman spectra at the \itO\itT phase transition can be understood on the basis of the interactions between the lattice and the MA molecules.
    As described in previous works~\cite{Ong2015,Lee2015b,Varadwaj2019}, there exists hydrogen bonding (H-bonds) interaction between MA$^+$ cations and the negatively charged [PbI$_3$]$^-$ cages. At low temperatures, each PbI$_6$ cage does not have enough thermal kinetic energy to overcome its H-bonding interactions, and thus PbI$_6$ rotation is a nearly-harmonic vibration (libration).
    Similarly, the movement of the MA molecule, either translation or rotation, is also restrained by the H-bonds.
    At higher temperatures, increased PbI$_6$ rotational energy allows the octahedra to dynamically break the H-bonding,  freeing the movement of the MA molecules.
    Fig.~\ref{Mode_trajectory}d shows that the MA molecules are allowed to visit more orientations in the  \itT-phase.
    The weight of the MA translational motion below 50~\wnws also increases in the \itT phase compared to the \itO phase (Fig. S9 in SM~\cite{SM}).
    Thus the \itO\itT phase transition marks the onset of orientational disorder of the MA$^+$ ion. In Fig.\ S11  in the SM~\cite{SM}, we present a cartoon of the octahedral cage for each phase to illustrate this special unlocking phenomenon.

	
    Having demonstrated the advent of anharmonicity in the \itT phase, we move on to explain its evolution with increasing temperature. 
    We decompose the low-frequency relaxational motion ($\approx$~1~\wn) into four different PbI$_6$ rotations (Fig. 3a and Fig. S13 in SM~\cite{SM}).
    The rotational motions are separated according to the rotation axis and relative direction.
    For example, z-out means the out-of-phase PbI$_6$ rotation along the \textit{z} axis. 
  Figures~\ref{Real_Space}b-d show the amplitude of the dynamic rotation for different phases.
  Table~S2 in SM~\cite{SM} tabulates the equilibrium octahedral tilting angles.
    Large amplitude rotations are indicative of strongly anharmonic, relaxational motions while a small amplitude is indicative of quasi-harmonic, vibrational motion.  
    For the \itT phase (Fig.~\ref{Real_Space}b-c) the amplitude of the dynamic rotation around the \textit{z}-axis is small.
    In contrast, the rotation amplitudes about the \textit{x} and \textit{y} axes are quite significant, revealing strong dynamic processes occurring along these two directions.
    The amplitudes grow even larger as temperature increases from 200~K to 300~K.
    This indicates that the relaxational rotation happens only around the two short axes of the tetragonal cell. Since in the \itT phase the amplitude of rotation around the \textit{z}-axis is small, the distortion in the \textit{xy} plane is not altered and thus the corresponding peak P3 remains sharp and distinct throughout this range of temperature.
    In the \itC phase (Fig.~\ref{Real_Space}d), however, PbI$_6$ rotations along all three axes are activated, and the equilibrium octahedral tilting angles for all three axes are close to zero ($a^0a^0a^0$).
    Interestingly, there is a temporal phase shift between the three out-of-phase rotations (see SM~\cite{SM}, Fig.\ S14), meaning that at one time, significant rotation occurs only about one axis while for the other two axes, the rotational amplitudes are quite small.
    This leads to instantaneous \itT phase structures formed along different directions.
    Therefore, apart from the abrupt structural rearrangements at the \itO\itT phase transition, the PbI$_6$ relaxational rotation is another important source of anharmonicity that increases continuously with temperature.
    Our  analysis clearly reflects the highly anharmonic behavior of the \itT phase. This contradicts the 0~K phonon dispersion computations which do not display negative frequencies~\cite{Yue2016,Ma2019}. 

In conclusion, we demonstrate strongly anharmonic structural dynamics and the presence of soft mode-like behavior in tetragonal MAPI.
We have shown how the atomic motions and phase sequence are dominated by temperature-activated relaxational motion of the PbI$_6$ octahedra and MA molecules.
The abrupt unlocking of the H-bonds at the \itO-\itT phase transition marks the onset of anharmonicity.
This enables the rotation of MA molecules and tilting in PbI$_6$ cages leading to large amplitude anharmonic motions.
The anharmonicity evolves with temperature due to the continuous damping of the PbI$ _{6} $ rotational modes throughout the \itT phase.
This also explains the liquid-like relaxational behavior observed in Raman spectra and reflects the breakdown of the phonon picture in HPs.
	
	\subsection{Acknowledgments}
		The authors would like to thank Dr. Tsachi Livneh
		(NRC) for fruitful discussions, Dr. Ishay Feldman (WIS)
		for performing X-Ray diffraction measurements and Dr.
		Lior Segev (WIS) for developing the Raman software.
		R. S. acknowledges FGS-WIS for financial support. O.
		Y. acknowledges funding from: ISF
		(1861/17), BSF (grant No.
		2016650) , ERC (850041 - ANHARMONIC), Benoziyo Endowment Fund, Ilse Katz Institute, Henry Chanoch
		Krenter Institute,
		Soref New Scientists Start up Fund, Carolito Stiftung,
		Abraham \& Sonia Rochlin Foundation, E. A. Drake and R. Drake and the Perlman Family. Z. D. and L. G.
		acknowledge support from the US National Science Foundation, under grant DMR-1719353. J. Z. acknowledges
		support from a VIEST Fellowship at the UPenn. A. M. R. acknowledges support from the
		Office of Naval Research under Grant N00014-17-1-2574.
		The authors acknowledge computational support from
		the High-Performance Computing Modernization Office.

	\newpage
	\widetext

	\begin{center}
		\large{\textbf{The tetragonal phase of CH$ _{3} $NH$ _{3} $PbI$ _{3} $ is strongly anharmonic }}
	\end{center}
	
	\begin{center}
		Rituraj Sharma$ ^{1*} $, Zhenbang Dai$ ^{2*} $, Lingyuan Gao$ ^{2} $, Thomas M. Brenner$ ^{1} $, Lena Yadgarov$ ^{1} $, Jiahao Zhang$ ^{2} $, Yevgeny Rakita$ ^{1} $, Roman Korobko$ ^{1} $, Andrew M. Rappe$ ^{2\dagger} $, Omer Yaffe$ ^{1\ddagger} $
	\end{center}
	
	\begin{center}
		\textit{$^1$Department of Materials and Interfaces, Weizmann Institute of Science, Rehovoth 76100, Israel\\
			$^2$Department of Chemistry, University of Pennsylvania, Philadelphia, Pennsylvania 19104, USA}
	\end{center}

	\setcounter{figure}{0}
	\setcounter{page}{1}
	\setcounter{equation}{0}
	\setcounter{table}{0}
	\setcounter{subsection}{0}
	
	\renewcommand{\thefigure}{S\arabic{figure}}
	\renewcommand{\thetable}{S\arabic{table}}
	\renewcommand{\theequation}{S\arabic{equation}}
	\renewcommand{\thesubsection}{\arabic{subsection}}
	\renewcommand{\bibnumfmt}[1]{[S#1]}
	\renewcommand{\citenumfont}[1]{S#1}
	\begin{center}
		\section{Supplemental Material}
	\end{center}

	\subsection*{Experimental methods}
	High quality, MAPI single crystals were grown at room temperature using the Anti-solvent method as discussed elsewhere~\cite{Rakita2017}. 
	The crystallographic orientation the crystal was determined using powder X-ray diffraction (XRD) (See Fig. S1). 
	Raman experiments were performed on the same face for which XRD was obtained.
	
	Low frequency micro-Raman scattering measurements were performed in a customized set-up fitted with a confocal microscope and optical cryostat (Janis, USA) evacuated to 10$^{-5}$~Torr.
	1064~nm line from an Nd:YAG laser was guided into a x50/0.42 NA, C-coated IR objective, using a 90/10 volume holographic grating (VHG) beam splitter (Ondax Inc., USA) in a backscattered configuration.
	The polarization of the beam was maintained using fixed Glan-Laser polarizers in the incident and scattered beam path.
	A motorized $\lambda$/2 wave plate controls the direction light polarization of with respect to the sample.
	An achromatic $\lambda$/2  wave plate rotates 0$^\circ$ or 45$^\circ$ relative to the polarizer to filter the scattered radiation parallel or cross-polarization directions.
	Two 90/10 VHG notch filters (each having OD $>$4 rejection with a spectral cut off $\pm$7~cm$ ^{-1} $ around 1064-nm) discard the Rayleigh-scattered light.
	The scattered beam is then imaged onto the entrance slit of a 1~m focal length spectrograph (Horiba FHR-1000), dispersed by a 950 lines/mm grating onto a liquid N$ _{2} $ cooled InGaAs detector (Symphony II, Horiba).

	\subsection*{Computational methods}
	
	We perform Born-Oppenheimer AIMD with the Quantum-ESPRESSO code. A $\sqrt{2}\times\sqrt{2}\times2$ supercell containing four formula units of MAPI is chosen. We use the PBE31 functional and the Tkatchenko-Scheffler scheme for dispersive interactions. A $3\times3\times2$ k-point grid is used, and the plane-wave cutoff energy is 50 Ry. The DFT total energy is converged to 10$^{-7}$ Ry/cell. Spin-orbit coupling is not taken into account, as it does not affect the structural properties of lead-halide perovskites. To compare with experiments, AIMD on MAPI was performed at 77 K, 200 K, 300 K, and 500 K, corresponding to orthorhombic, tetragonal, and cubic phases. We choose 500 K for the simulation of cubic phase to make it more distinguishable from the tetragonal phase. This temperature is well above the transition temperature, and the perovskite structure can be maintained well in our AIMD simulation. The time step of AIMD is 1 fs, and we equilibrate the systems with a velocity-rescaling thermostat (rescale every 10 steps) for 10 ps. We run another 30 ps to obtain a converged autocorrelation function. To obtain the Raman spectra, density functional perturbation theory (DFPT) with a denser $6\times6\times4$ k-point grid is used to calculate the polarizability at a time interval of 100 fs.
	
	We employ the autocorrelation function approach to compute Raman activities where we relate the Raman tensor to the dynamical autocorrelations of the polarizability tensor to address the anharmonic effects. Peak intensities of the non-resonant Stokes Raman spectra $I^\mathrm{\nu}$ can be computed as \cite{Putrino2002}: 
	\begin{equation}
		I^{\mathrm{\nu }}_{ij} \mathrm{\propto } \ {\mathrm{\omega }}^{\mathrm{2}}_{\mathrm{\nu }}\int{{\left\langle {\mathrm{\alpha }}_{ij}\left(\mathrm{\tau }\right){\mathrm{\alpha }}_{ij}\left(t+\mathrm{\tau }\right)\right\rangle }_{\tau }e^{-i{\mathrm{\omega }}_{\mathrm{\nu }}t}dt}
	\end{equation}
	Here, $i (j)$ represents the polarization component of the incident (scattered) light, $\omega_{\nu}$ is the frequency of the phonon mode $\mathrm{\nu }$, and ${\mathrm{\alpha }}_{ij}$ is the polarizability tensor. The polarizabilities of selected structures are obtained from the ab-initio molecular dynamics (AIMD) trajectory, and we employ the Wiener-Khintchine’s theorem to calculate the autocorrelation functions. 
	
	We calculate frequency-filtered trajectories by first performing a Fourier transform on the real-space MD trajectories (from \textit{t} to $\omega$), and then filter the $\omega$-space trajectories within a frequency window$\mathrm{\ }\mathrm{\Delta }\mathrm{\omega }$, and finally perform an inverse Fourier transform back to \textit{t}. The whole procedure is expressed as:\\
	
	\noindent ${\mathop{R}\limits^{\rightharpoonup}}_{\mathrm{filtered}}\left(t;{\mathrm{\omega }}_0\right)$=
	\begin{equation}
		{\mathcal{F}}^{-1}\left[\mathrm{\Theta }\left(\mathrm{\omega }-({\mathrm{\omega }}_0-\mathrm{\Delta }\mathrm{\omega })\right)\mathrm{\Theta }\left(-\mathrm{\omega }+({\mathrm{\omega }}_0\mathrm{+}\mathrm{\Delta }\mathrm{\omega })\right)\mathcal{F}\left[\mathop{R}\limits^{\rightharpoonup}(t)\right]\right]
	\end{equation}
	
	In addition, we establish a basis of modes that can fully describe atomic motions, including Pb-I stretching, PbI$_6$ distortion, PbI$_6$ rotation, Pb off-center motion, and MA molecular translation (we take the molecule as a unit). We project the frequency-filtered trajectories onto this basis via the following equation:
	\begin{equation}
		W_{\mathop{u}\limits^{\rightharpoonup},{\omega }_0} = \frac{1}{N_t}\sum_t\frac{\vert {\mathop{R}\limits^{\rightharpoonup}}_{filtered}(t;\omega_0)\cdot\mathop{u}\limits^{\rightharpoonup}\vert^2}{\vert\mathop{R}\limits^{\rightharpoonup}\vert^2(t;\omega_0)}
	\end{equation}
	With the weight $W_{\mathop{u}\limits^{\rightharpoonup},{\omega }_0}$ of each component, we can tell the specific motions corresponding to the Raman spectral weight at frequency $\omega_0$.

	\noindent \subsection*{Fitting Procedure}
	\noindent We have used the imaginary part of the damped Lorentz oscillator model to simulate the Raman scattering data. The experimentally measured Raman scattering spectrum is expressed as eq.~\ref{s1}
	\begin{eqnarray}
		\label{s1}
		&I_{exp}(\nu,\nu_i,\Gamma_i)=c_{BE}(\nu)\left(\sum_{i=1}^{n} c_i\frac{\left| \nu\right| \left| \nu_i\right| \Gamma_i^2}{\nu^2\Gamma_i^2 + (\nu^2 - \nu_i^2)^2}\right)
	\end{eqnarray}
	where $ \nu $ is the spectral shift, $ \nu_i $ and $\Gamma_i $ are respectively the resonance energy and damping coefficient of the Lorentz oscillators and $ c_{i} $ are unitless fitting parameters for the intensities of the Lorentz oscillator components. The spectral shift $ \nu $, the parameters  $ \nu_i $ and $\Gamma_i $ are in wavenumber units. The approximate lifetime of the phonons can be calculated as 
	$\tau$=$ \dfrac{1}{2\pi c \Gamma_i} $, where c is the speed of light. The prefactor $ c_{BE}(\nu) $ accounts for the thermal population from the Bose-Einstein distribution. The Stokes signal is proportional to 1+$ \mathit{n} $, and the anti-Stokes signal is proportional to \textit{n}, with \textit{n} being the Bose-Einstein distribution. The Bose-Einstein prefactor is written as eq. S2
	\begin{equation*}
		c_{BE}(\nu) = \begin{cases}
			n+1 & \quad \nu_0 \ge  0, \\
			n & \quad \nu_0 < 0.
		\end{cases}
	\end{equation*}
	
	We remove the spectral artifacts from the notch filter around 0 cm$^{-1}$, and fitted the Stokes scattering with equation S1 using fitting parameters $ c_{i} $; $\nu_i$ and $ \Gamma_i$. Numerical fitting was carried out in the customized MATLAB code and Igor Pro 8. 
	
	To fit the parallel polarization orientation (PO) data of the tetragonal phase, we have identified six peaks. Peaks P1 and P3 appear only angle 140$ ^\circ $ and P2 appears only at 50$ ^\circ $. Peaks P4, P5 and P6 remain common for all PO angles. The data at both 50$ ^\circ $ and 140$ ^\circ $ angles are separately fitted to find the positions and widths of the peaks. Thereafter, the obtained parameters were fixed to fit the entire PO dependence, keeping only the intensity of modes ($ c_i $) as variable. To fit the higher temperature, cubic data, the Raman modes are further softened and damped to produce a good overall fit containing the central peak. Then the same technique was applied to fitting the intensities of the PO as in the tetragonal phase. The same parameters for five peaks (P1, P3, P4, P5 and P6) were fixed to fit the cross data. As P2 was not visible in cross configuration, it was not used to fit cross data. The $ \Gamma_i $ values obtained from fitting were used to calculate the lifetime of the Raman modes.
	
	\newpage
	
	\begin{table}[ht!]
		\caption{Fit position and width parameters from fits to the polarization dependent Raman data sets for each temperature measured. These parameters were held constant for all polarization angles and both analyzer configurations, except for P2, which is only present in the parallel configuration.}	
		\begin{ruledtabular}	
			\begin{tabular}{ccccccccc}

				&\multicolumn{2}{c}{200 K}&	\multicolumn{2}{c}{250 K}&\multicolumn{2}{c}{300 K}&\multicolumn{2}{c}{360 K}\\
				\hline
				Peak label&	$\omega$ (cm$^{-1}$)&$\Gamma$ (cm$^{-1}$)&	$\omega$ (cm$^{-1}$)&$\Gamma$(cm$^{-1}$)&$\omega$ (cm$^{-1}$)&$\Gamma$ (cm$^{-1}$)&$\omega$ (cm$^{-1}$)&$\Gamma$ (cm$^{-1}$)\\
				\hline
				P1&	33.0&	19.2&	31.9&    23.3&    30.5&   29.0&    23.0&   24.0\\
				P2&	37.5&	25.0&	37.0&    29.0&    36.5&   38.0&    -&   -\\
				P3&	45.7&	6.2&	44.8&    7.8&    44.1&   10.5&    44.0&  33.0\\
				P4&	66.0&	24.0&	68.5&    35.0&    67.1&   44.0&    64.0&  40.0\\
				P5&	93.0&	34.2&	95.0&    30.0&    98.0&   32.9&   100.0 & 45.0\\
				P6&	126.1&	46.4&	128.5&    51.5&    129.8&   51.5&   132.5 &  56.2 \\
				
			\end{tabular}
		\end{ruledtabular}
		\label{Table1}
	\end{table}

	\begin{table}[ht!]
		\caption{The equilibrium octahedral tilting angles for \textit{t}- and \textit{c}-phases along the x, y and z directions}	
		\begin{ruledtabular}	
			\begin{tabular}{cccc}
				Phase&\multicolumn{3}{c}{Equilibrium octahedral tilting angle}\\
				&x&y&z\\
				\hline
				Tetragonal (200 K)&3.65&2.18&24.66\\
				Tetragonal (300 K)&7.55&4.49&24.20\\
				Cubic&4.94&1.66&2.96\\
			\end{tabular}
		\end{ruledtabular}
		\label{Table2}
	\end{table}	
	
\pagebreak
	
	\begin{figure}[ht!]
		\includegraphics[width=3.4in]{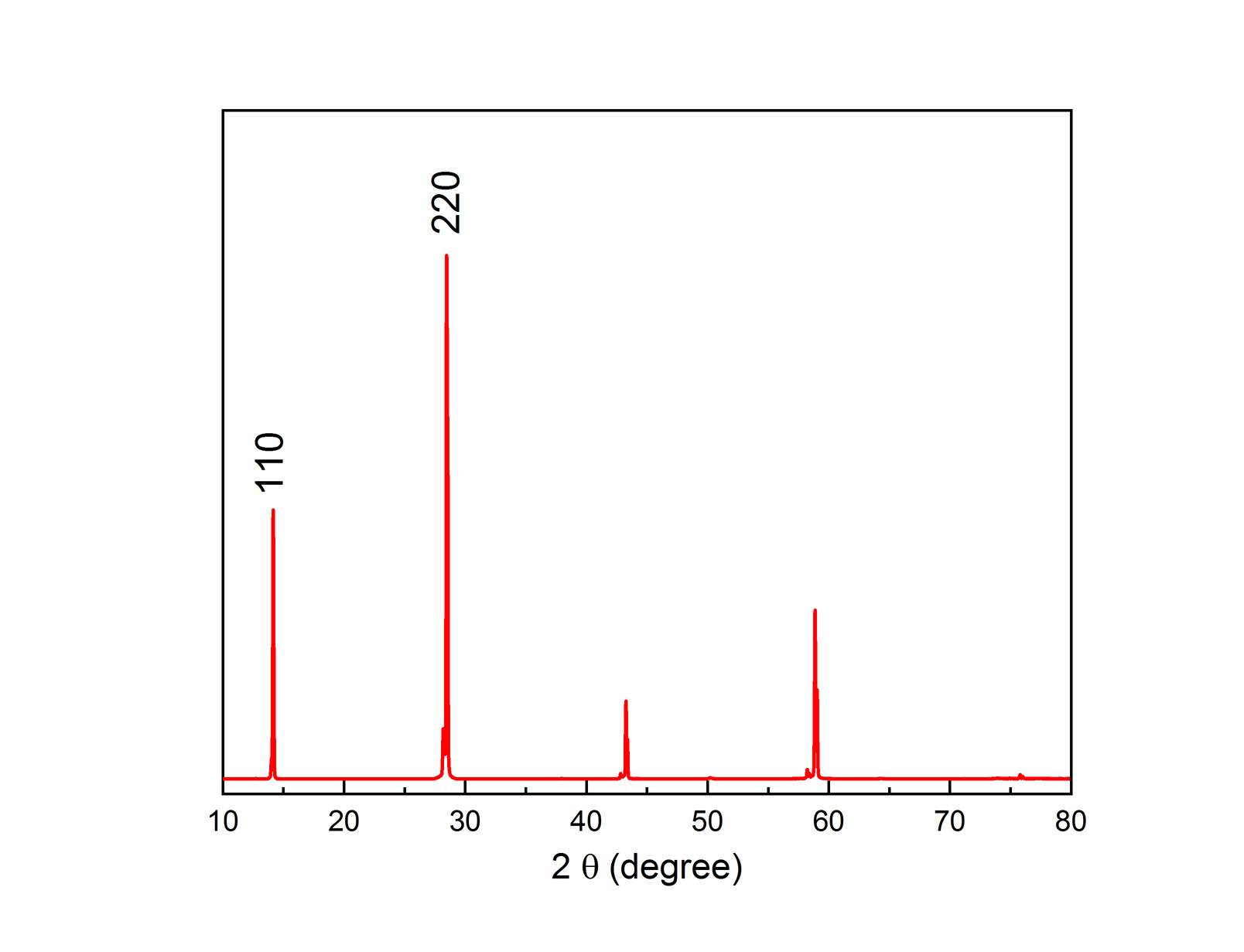}
		\caption{Room temperature XRD of MAPI single crystal.}
	\end{figure}
\newpage
	
	\begin{figure}[ht!]
		\includegraphics[width=6.6in]{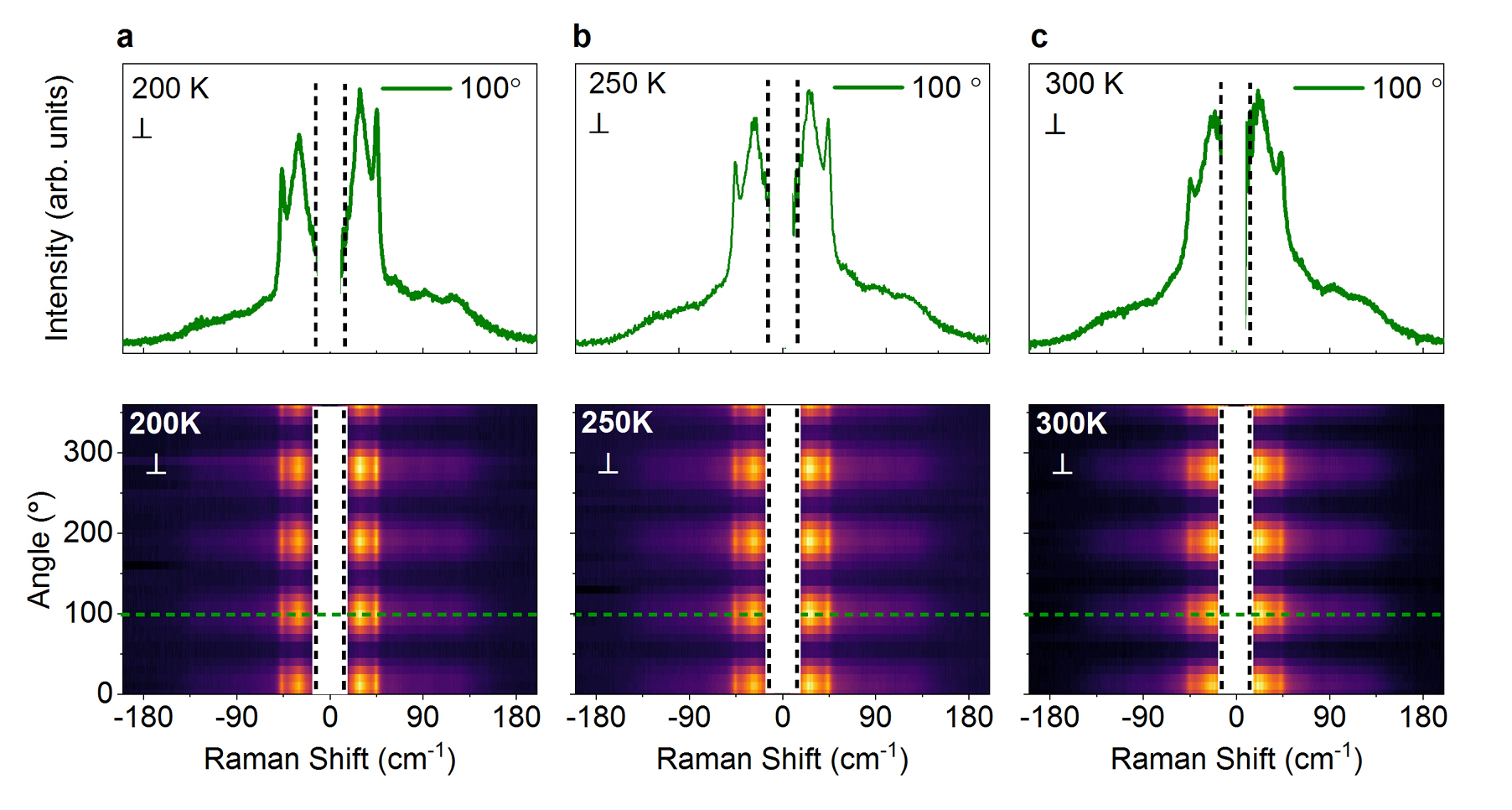}
		\caption{Contour plots for PO dependence of the Raman spectra of MAPI in cross configuration for (a) 200 K, (b) 250 K and (c) 300 K. The 0$^\circ$ polarization angle is arbitrarily defined and is identical for all measurements. Cross sections of the contours corresponding to the angular positions shown by dotted lines are also shown at the top of each contour. Strong polarization angle dependent modulation of the Raman intensity is observed with 90$^\circ$ period.}
	\end{figure}
	
	\begin{figure}[ht!]
		\includegraphics[width=6in]{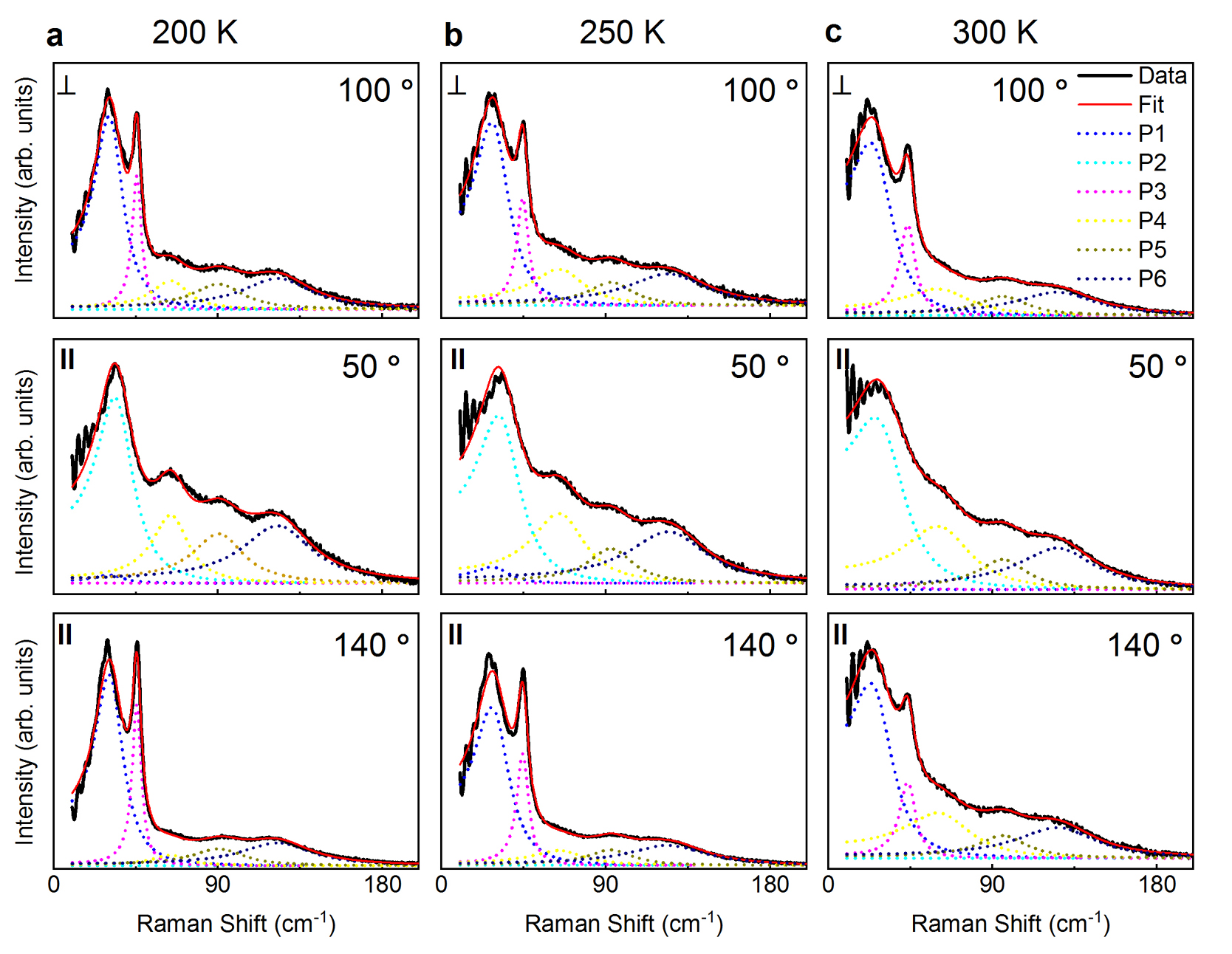}
		\caption{Fitting of tetragonal phase Raman spectra using a multi Lorentz oscillator. The experimental data are fitted to six Lorentz at temperatures 200 K (a), 250 K (b) and 300 K (c) using equation S1.  }
	\end{figure}

	\begin{figure}[!ht]
		\includegraphics[width=3in]{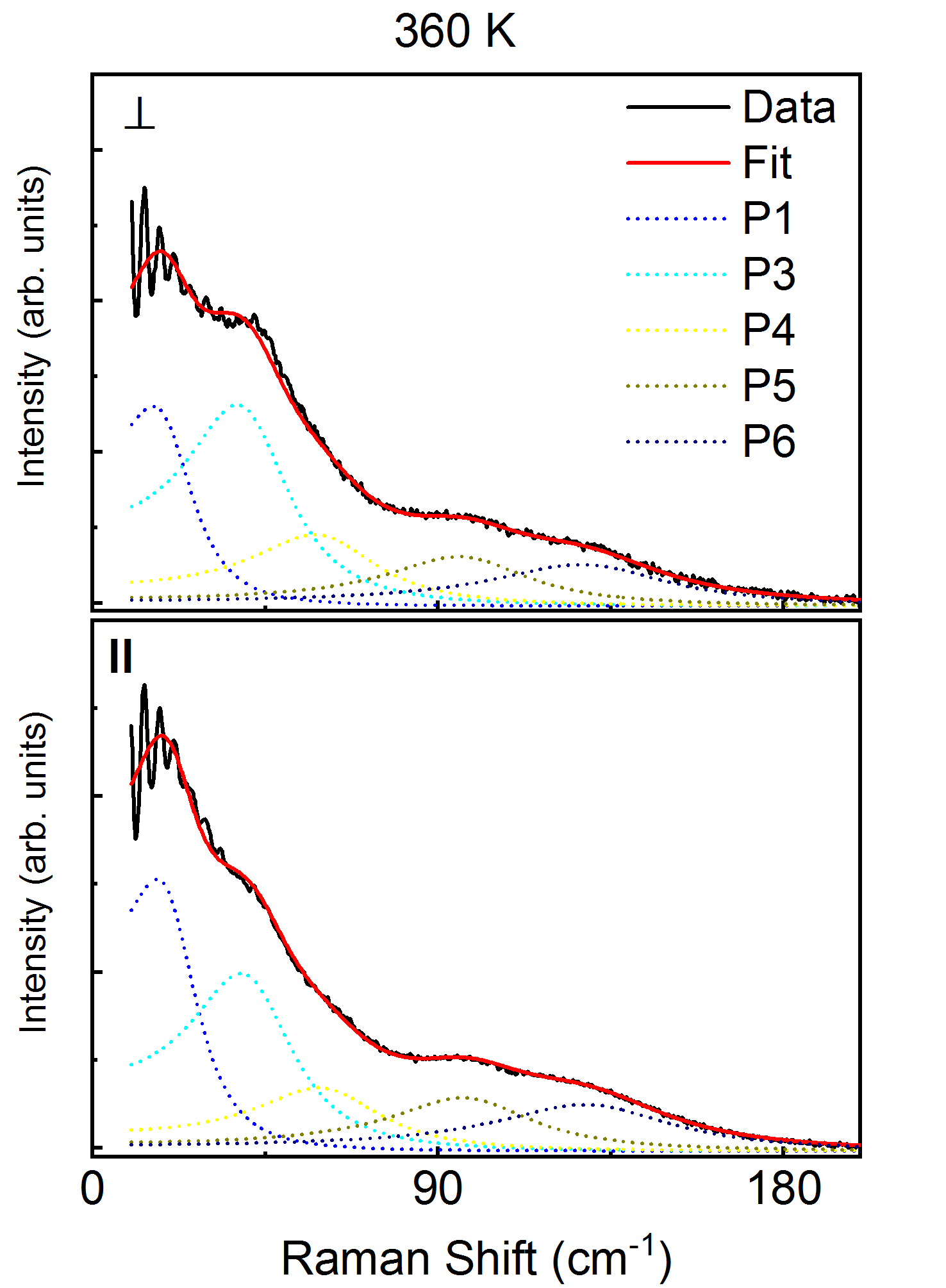}\caption{Fitting of cubic phase Raman spectra using multi Lorentz oscillator at 360 K using equation S1.  }
	\end{figure}

	\begin{figure}[ht!]
		\includegraphics[width=6in]{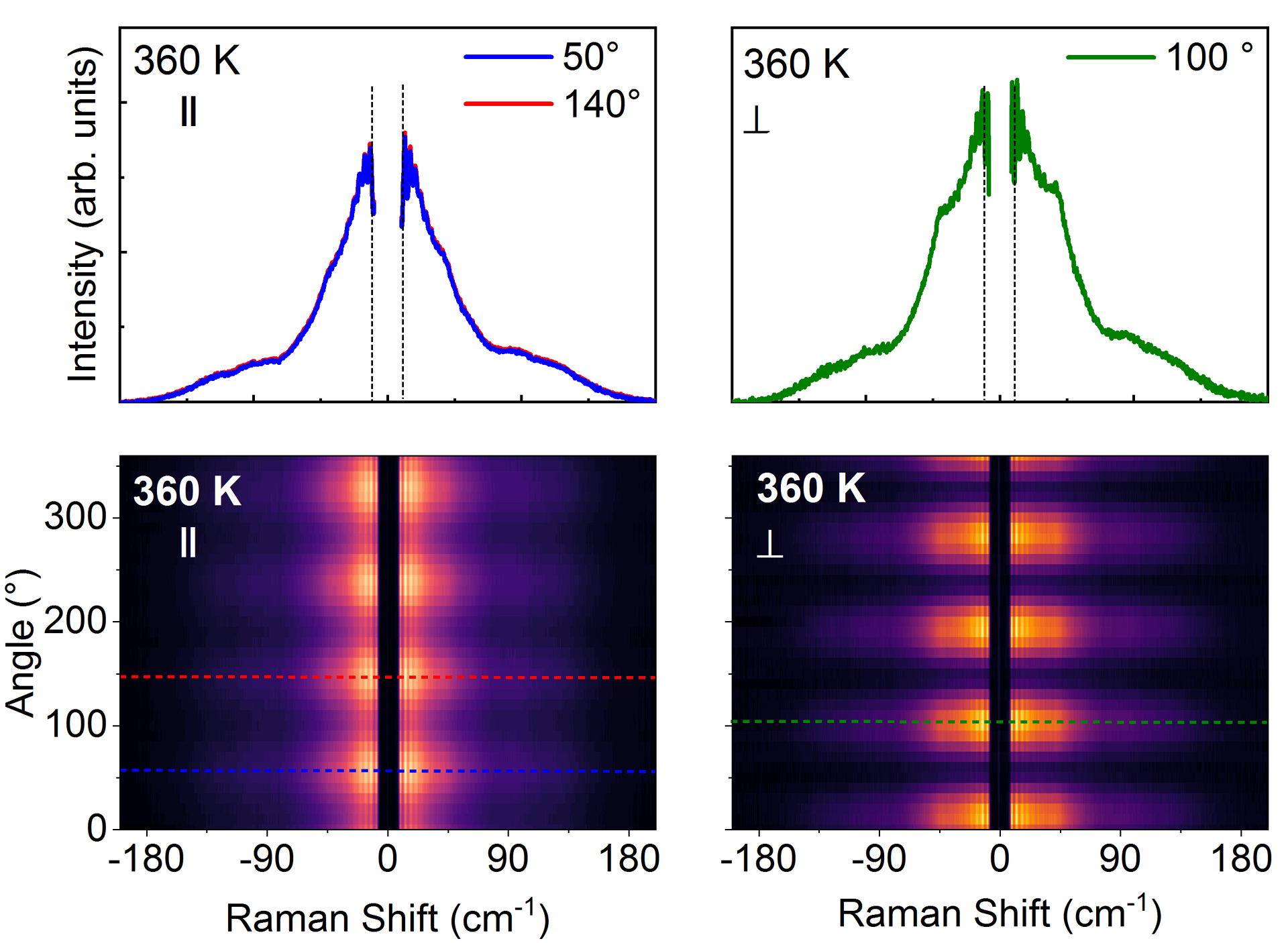}
		\caption{PO dependence of cubic phase (360 K) Raman spectra in (a) parallel and (b) cross configurations. 180 $ ^{\circ} $ periodicity is observed in both configurations. Line shape at 50 $ ^{\circ} $ and 150 $ ^{\circ} $ PO angles are same in parallel configuration representing the isotropy of the cubic structure.}
	\end{figure}

	\begin{figure}[ht!]
		\includegraphics[width=3.5in]{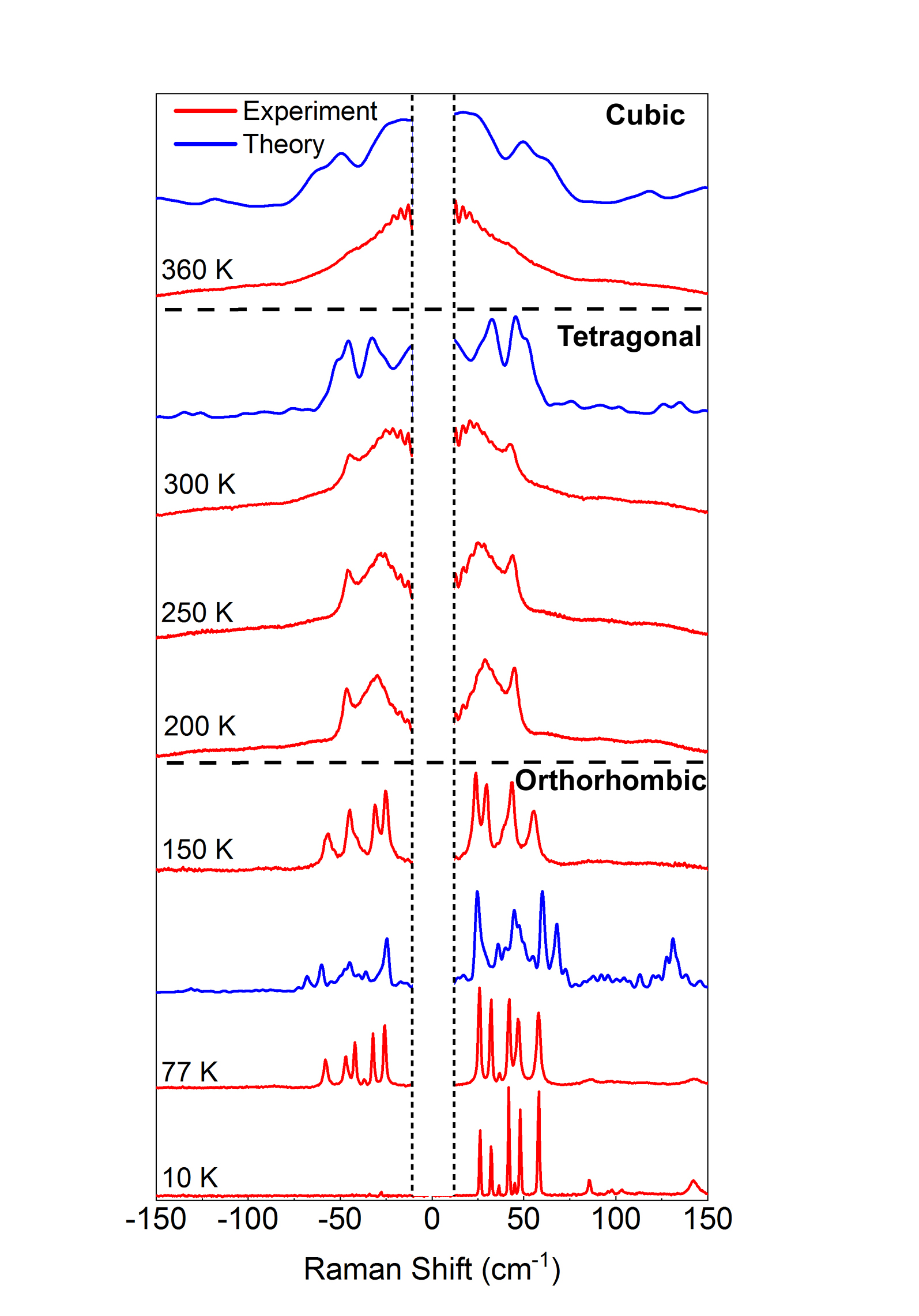}
		\caption{Experimental and theoretical low-frequency Raman spectra 
			of MAPI at temperatures across the \textit{o-}, \textit{t-} and \textit{c-} phase sequences from 10~K to 360~K (experiment=red; theory=blue).
			The phases are separated by horizontal dotted lines.
			The \textit{o-t} phase transition is characterized by evolution of sharp peaks to an envelope of two broad peaks.
			Calculated Raman spectra for each phase are stacked over experimental spectra.
			The theoretical spectra are calculated at 77~K (orthorhombic), 300~K (tetragonal), and 500~K (cubic) using MD simulations.
			The spectral region 0 $ \pm $ 10 cm$^{-1}$ (marked by the vertical dashed lines) is masked by the notch filter and has been omitted.}
	\end{figure}
	
	\newpage
	\begin{figure}[ht!]
		\includegraphics[width=6in]{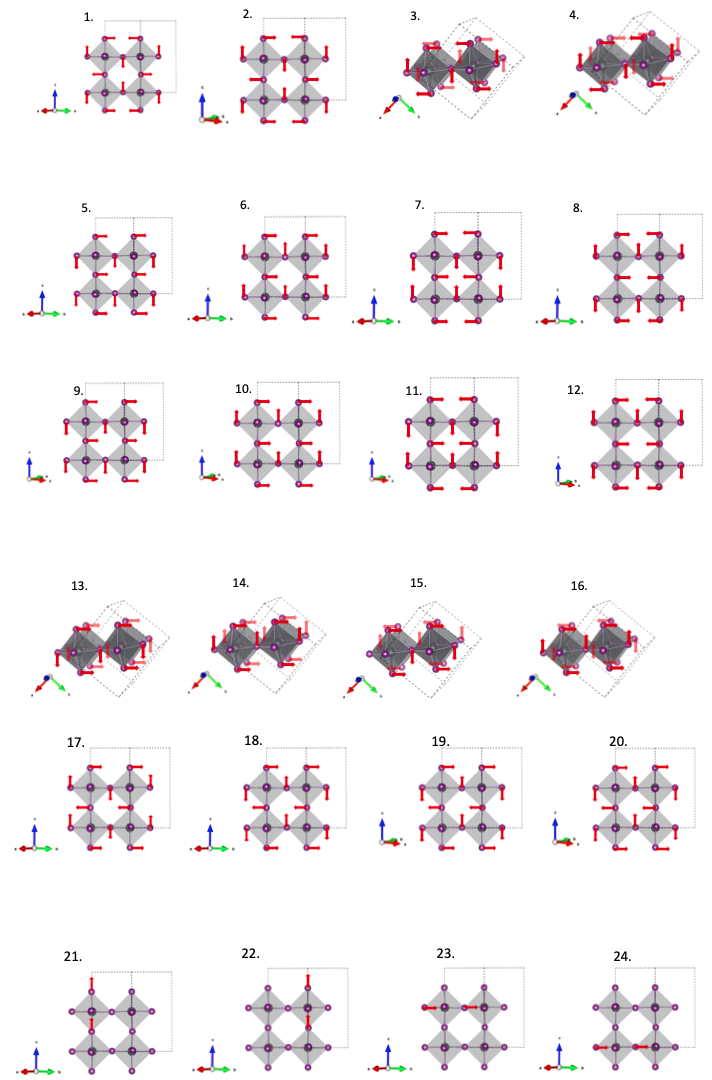}
	\end{figure}
	\begin{figure}[ht!]
		\includegraphics[width=6in]{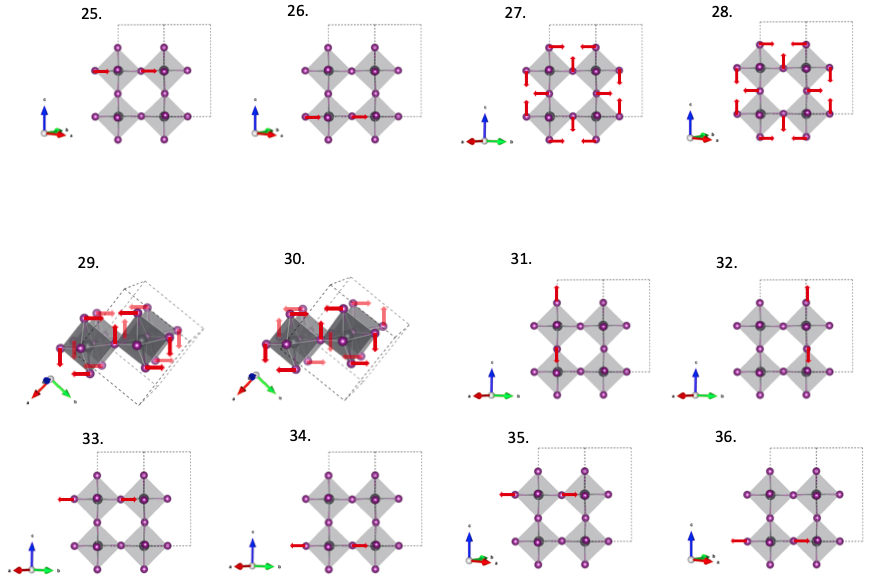}
		\caption{Complete definition lattice modes. The preliminary classifications shown in the legend of Figure. 2 are as follow: mode 1, 2, 3, and 4 are belong to “Rotation”, mode 5-30 are belong to “Distortion”, and mode 31, 32, 33, 34, 35, and 36 are belong to “Stretching”. Here, we can see that mode 27 and 28 are equivalent, and they have the same distortion pattern as mode 30.}
	\end{figure}

	\pagebreak
	\begin{figure}[ht!]
		\includegraphics[width=3.4in]{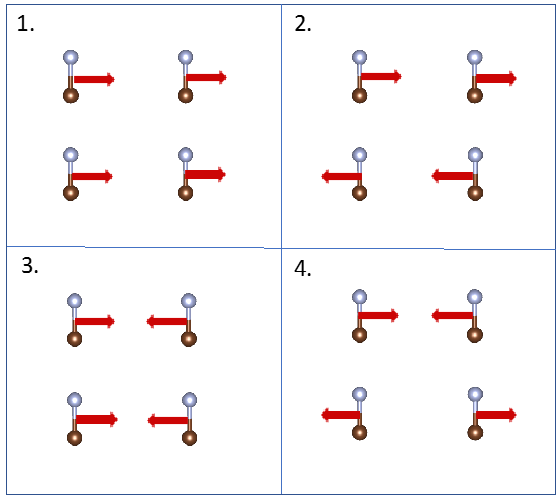}
		\caption{Illustration of the 4 MA molecular translation modes (H atoms are omitted for clarity). Similar patterns can be defined for other directions.}
	\end{figure}
	
	\newpage

	\begin{figure}
		\includegraphics[width=6in]{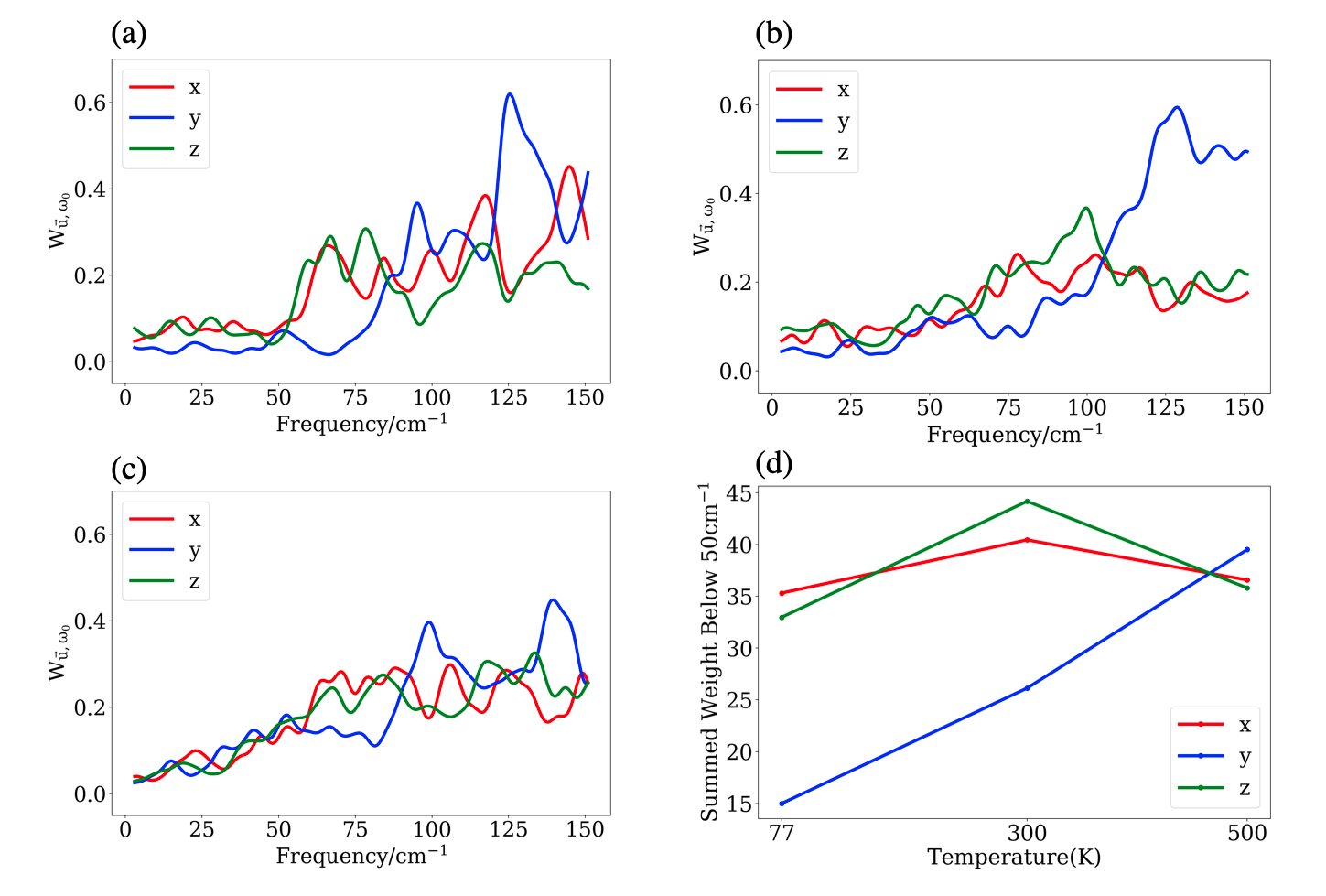}
		\caption{Mode projection for molecular translations along different directions for 77 K (a), 300 K (b) and 500 K (c). (d) Summation of projections below 50 cm$ ^{-1} $ for different temperatures. It can be seen that at 77 K, the molecular translation for the low-frequency range (0 - 50cm$ ^{-1} $) in every direction has the lowest weight because of the “locking” of hydrogen bond formed between MA and [PbI$_6$] lattice. At tetragonal phase (300 K), this coupling is “unlocked” such that the weight of molecular translation increases. In cubic phase (500 K), since PbI$ _{6} $ rotation totally dominates the low-frequency range, the relative weight of overall molecular translation has a slightly decrease. But we also notice the consistent increase in y axis from t- to c-phase. This is because lattice constant is the shortest in tetragonal phase for pseudo-cubic cell (not the simulation cell), which sterically restrict its motion at tetragonal phase. When transitioning into cubic phase, all the three axes are equivalent so that the weight of y axis can continue increasing.}
	\end{figure}

	\begin{figure}[ht!]
		\includegraphics[width=6.5in]{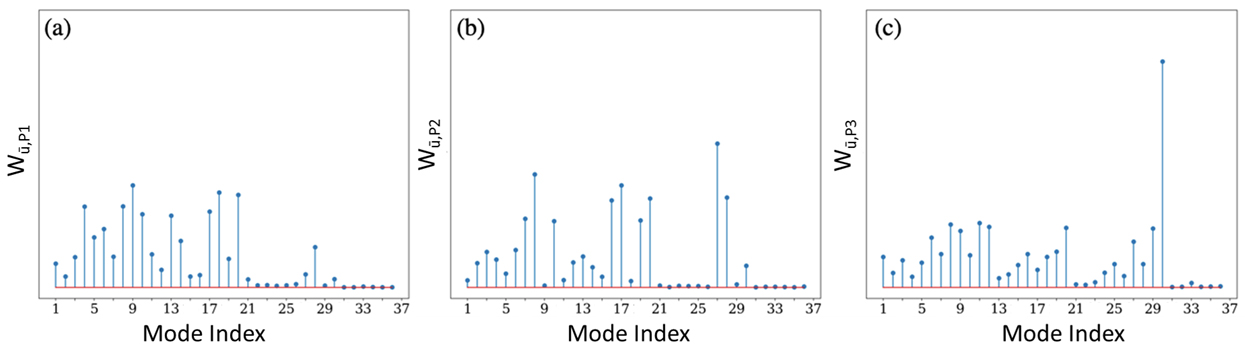}
		\caption{Assignment of low frequency peaks in PO measurement. The frequency window is 0.5cm$^{-1}$ in order to distinguish the closely lying peaks. The peak at $\approx$45 cm$^{-1}$ (P3) is mostly represented by mode 30, which is a symmetric distortion mode in xy plane. The hidden mode at $\approx$30 cm$^{-1}$ (P2), is mostly represented by mode 27 and mode 28, which are two equivalently symmetric distortions in different planes (pseudocubic xz, and yz planes). The distortion patterns of both the peaks are the same but in different planes. For the peak at $\approx$28 cm$^{-1}$ (P1), the motions are even more convoluted, so it’s hard to identify a specific motion for it. However, if we look at the modes 5-16 which have relative large weight, we can find that they all fall into the same pattern where each octahedron is shrinking on one side and expanding on the other side, so these kinds of asymmetric distortion could contribute to P2. On the other hand, modes 17-20 are of large weight, and their motions can be classified as PbI$_6$ rotation for two octahedrons and symmetric distortion for the other two octahedrons in the supercell. Therefore, this kind of motion could be another source.}
	\end{figure}

	\newpage

	\begin{figure}[ht!]
		\includegraphics[width=6in]{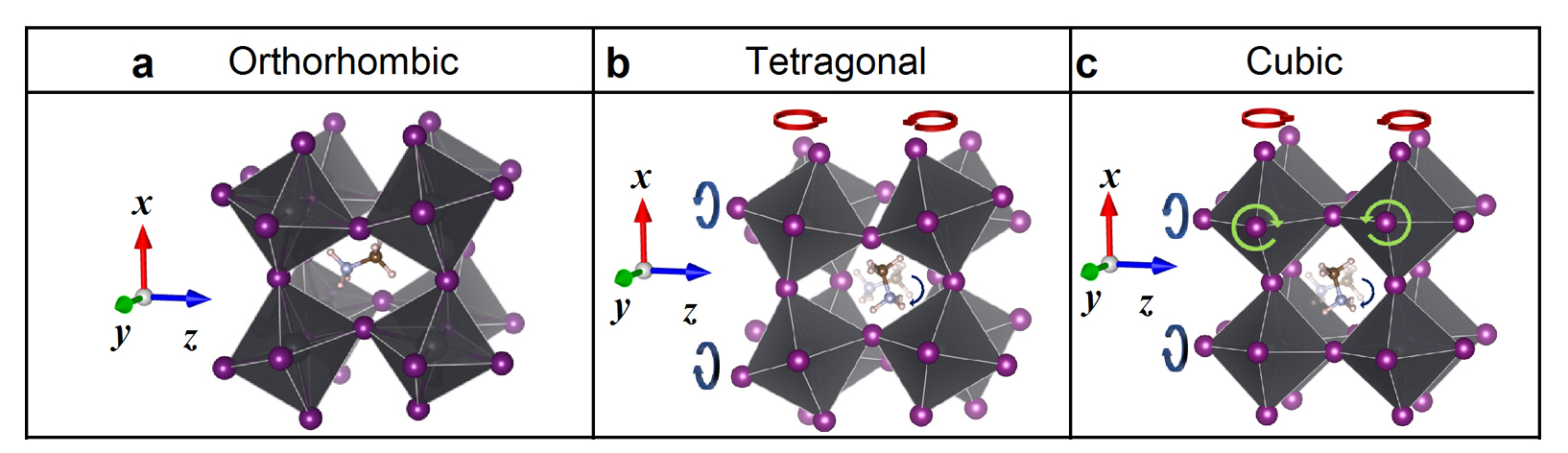}
		\caption{(a-b) Illustration of the unlocking of PbI$_6$ octahedra and MA molecule. In the \textit{o-} phase, the cage motions are restrained by H-bonds.
			The \textit{o-t} phase transition occurs when the H-bond break, leading to relaxaional rotations of the cage along certain directions (shown by circular arrows).
			In the \textit{c-} phase (c), all-axes PbI$_6$ rotations are activated.}
	\end{figure}

	\begin{figure}[ht!]
		\includegraphics[width=3in]{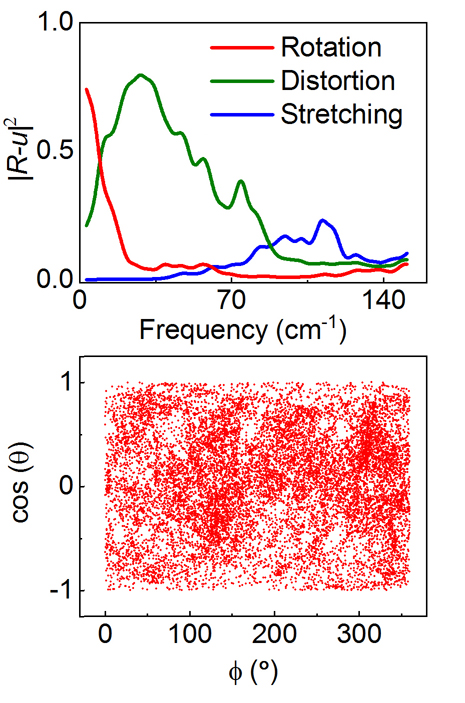}
		\caption{The projections of frequency-filtered MD for \textit{c-}phase.	A broad central feature for rotational modes indicates that relaxational motion dominates all other types of motions. The lower panel: Statistics of molecular orientations, showing complete randomness in the \textit{c-}phase. $\phi$ is the azimuthal and $\theta$ is the polar angle. }
	\end{figure}

	\begin{figure}[ht!]
		\includegraphics[width=3in]{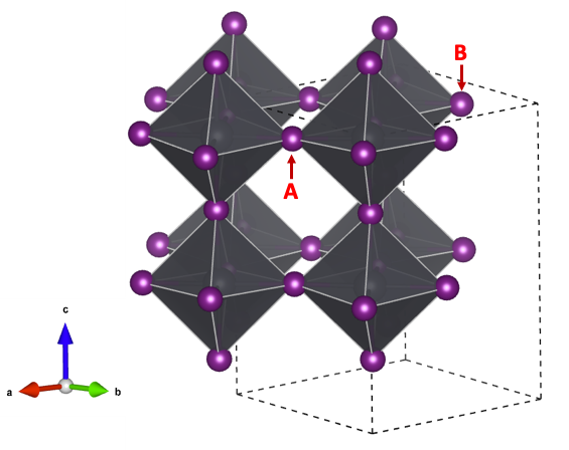}
		\caption{Illustration of the possible PbI$_6$ rotation modes in our simulated supercell. In general, PbI$_6$ rotation can happen along x, y, and z axis, and between two layers, the rotation can be either in-phase or out-of-phase. However, due to the size limitation of our simulated cell, only out-of-phase rotation can happen along the pseudo-cubic x and y axis, whereas both in-phase and out-of-phase rotation can occur along z axis.  The possible two types of rotation about z axis can be understood easily since there are two independent layers along z-axis for our supercell. However, for the pseudo-cubic x and y axis, which are the diagonals for a $ \sqrt{2}$ x $ \sqrt{2}$ x 2 supercell, the two-layers are interrelated. For example, A and B iodine atoms shown here are the images of each other. If atom A moves downward (-z), the B also has to move downward (-z). The similar argument can apply to other iodine atoms involved in a PbI$ _{6} $ rotation mode. Thus, we can see that the only possible PbI$ _{6} $ rotation mode along pseudo-cubic x and y axis is out-of-phase rotation.}
	\end{figure}

	\begin{figure}[ht!]
		\includegraphics[width=3in]{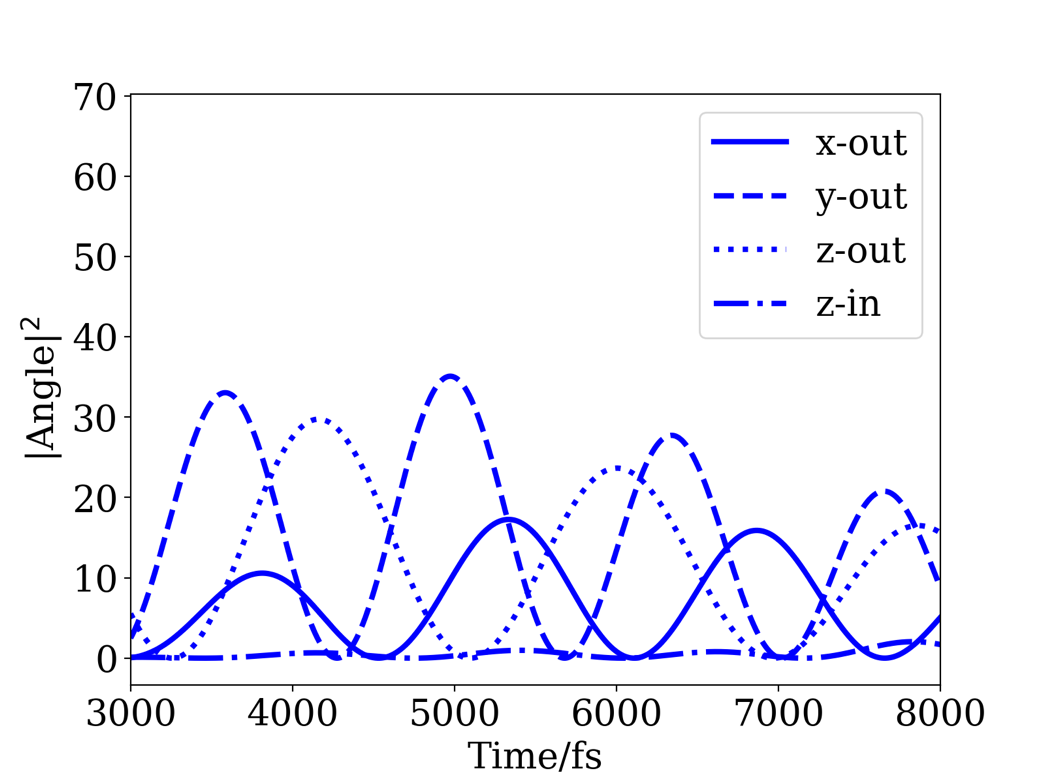}
		\caption{ The fluctuation of the square of PbI$ _{6} $ rotation in a randomly selected time domain. All the equilibrium tilting angles have been subtracted for clarity.  The temporal phase shift between the three out-of-phase rotations indicates that at one time, the significant rotation occurs only along one axis while for the other two axes, rotational amplitudes are quite small. This leads to instantaneous, tetragonal structures formed along different directions, and it shows the observed cubic phase is indeed the dynamical sampling of \textit{t}-phase along different orientations. }
	\end{figure}

\newpage

\end{document}